\documentclass[aps,prl,preprint,groupedaddress]{revtex4}
\usepackage{graphicx}
\begin{document}

%\preprint{}

\title{Mathematical Model of Hit Phenomena}


\author{Akira Ishii, Sanae Umemura, Takefumi Hayashi, Naoya Matsuda}
\email[Akira Ishii : ]{ishii@damp.tottori-u.ac.jp}
%\thanks{}
\affiliation{Department of Applied Mathematics and Physics, Tottori University, Koyama, Tottori 680-8552, Japan}

\author{Takeshi Nakagawa}
%\thanks{}
\affiliation{Dentsu Inc., 1 -8-1, Higashi-shimbashi, Minato-ku, Tokyo 105-7001, Japan}

\author{Hisashi Arakaki, Narihiko Yoshida}
%\thanks{}
\affiliation{Digital Hollywood University, DH2001 Bldg. 2-3 Surugadai, Kanda, Chiyoda-ku, Tokyo, 101-0062}

\begin{abstract}

The mathematical model for hit phenomena in entertainments is presented as a nonlinear, dynamical and non-equilibrium phenomena. The purchase intention for each person is introduced and direct and indirect communications are expressed as two-body and three-body interaction in our model. The mathematical model is expressed as coupled nonlinear differential equations. The important factor in the model is the decay time of rumor for the hit. The calculated results agree very well with revenues of recent 25 movies.


\end{abstract}

\maketitle

\newpage

\section{Introduction}

Hit phenomena in entertainments are very popular dynamical phenomena in economics. In contrast to the usual economic activities, there are no balances of demand and supply in hit phenomena. Thus, for analysis of hit phenomena, we consider it as time-dependent non-equilibrium phenomena. Moreover, especially for movie hits, the marketing share is not important. The significant factors for movie hit are the movie itself is attractive or not. Therefore, we can consider the hit for each movie individually. 

 In this study, we consider the hit phenomena of movies with both experimental and theoretical ways. For experimental viewpoint, we observe daily revenue data and daily blog-writing data for many movies. We also obtain the advertisement cost for the movies. For theoretical viewpoint, we present here a mathematical model for hit phenomena as non-equilibrium, nonlinear and dynamical phenomena. Comparing the simulated revenue with the observed revenue for the movies checks the model. The model presented in this paper is based on the previous works.\cite{ishii2005},\cite{ishii2007}
 
\section{Observation of hit evidence}

  We observe three data of recent 25 popular movies for daily revenue, daily advertisement costs and the entry number of blog writing in Japanese market. The revenue data is obtained from Nikkan Kogyo Tsushinsya in Japan. The advertisement cost data are obtained from Dentsu inc. The blog writing data is collected using the site "Kizasi". 
  
  First, as shown in fig.1, we found that the revenues decrease almost exponentially. This evidence is very natural, because the number of audience decrease monotonically due to the effect that person who watched the movie does not watch the same movie again. In fig.1, only the data E shows the sudden decrease. It happens due to the sudden accident of the actress of the movie. From the data, we found that the decay factors for most of all movies in fig.1 seems to be similar. The decay factor of the exponential decay is nearly 0.06 per day. It agrees well with the empirical rule of the Japanese movie market that the number of audience decrease roughly 6 percents.
  
\begin{figure}[htbp]
\begin{center}
\includegraphics[width=12cm,bb=0 0 667 471]{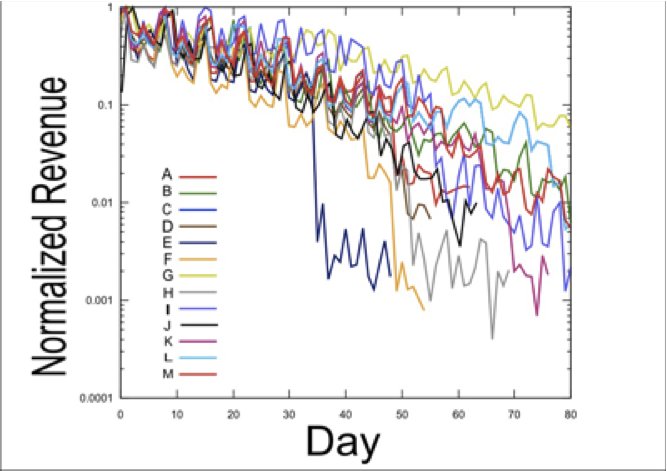}
\caption{Revenues of the movies are shown as functions of date. Revenue data is the value in Japanese market. The movie titles are as follows. A:Death Note, B:Death Note the last name, C: zoku Always, D: Close zero, E: Closed Note, F:Biohazard 3, G:Bushi no Ichibun, H:Ratatouille, I:fra-girl, J:Dainihonjin, K:Maiko haaaan!!, L:Koizora, M:HERO.}
\label{fig:fig1}
\end{center}
\end{figure}

  This exponential decrease can be explained easily by using a simple mathematical model. First, we define that the number of potential audience to be $N_0$ and the number of integrated audience at the time t to be $N(t)$. Thus, the number of people who have interest on the movie is $N_0-N(t)$. When we assume that the probability to watch the movie per one day is $a$, we obtain the equation to describe the number of audience to be 
\begin{equation} 
\label{eq:eq1}
\frac{dN(t)}{dt} = a(N_0 - N(t))
\end{equation}
The solution of the equation is
\begin{equation}
N(t) = N_0 - Ce^{-at}
\end{equation}
Thus, when the initial condition is $N(t)=0$ at $t=0$, we obtain
\begin{equation}
N(t) = N_0(1-e^{-at})
\end{equation}
It is clear that the result can be explain roughly the exponential decay of the audience shown in fig.1.

Next, we compare the daily change of the revenue data with the number of the entry of the blog writing. The typical results are shown in figs.2-5. The data shows the exponential decay. The quasi-periodic enhancement corresponds to the weekend effect. The revenue of each movies increases at Saturday and Sunday. The weak peaks in the middle of the week correspond to the ladies day discount system at Wednesday that is very popular in Japan.  From these figures, we found that the daily change of the revenue and the blog writing is very similar. This feature is found for all 25 movies. It means that people write their opinion on their blog with a certain probability. Figs2-3 shows us that the probability is almost constant.

According to the similarity, we propose here that the number of blog-writing entry can be use as {\it quasi-revenue} for each movie. This quasi-revenue is very useful, because the quasi-revenue can be defined even before the release day. Thus, we can catch the increasing purchase intention signal using the quasi-revenue.

\begin{figure}[htbp]
\begin{center}
\includegraphics[width=12cm,bb=0 0 300 240]{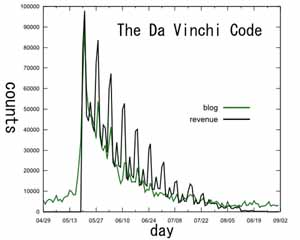}
\caption{The daily data of revenue and the blog-writing entry number for the Da Vinci Code in Japanese market. The blog data are normalized to adjust the peak of the revenue at the release day.}
\label{fig:fig2}
\end{center}
\end{figure} 

\begin{figure}[htbp]
\begin{center}
\includegraphics[width=12cm,bb=0 0 834 667]{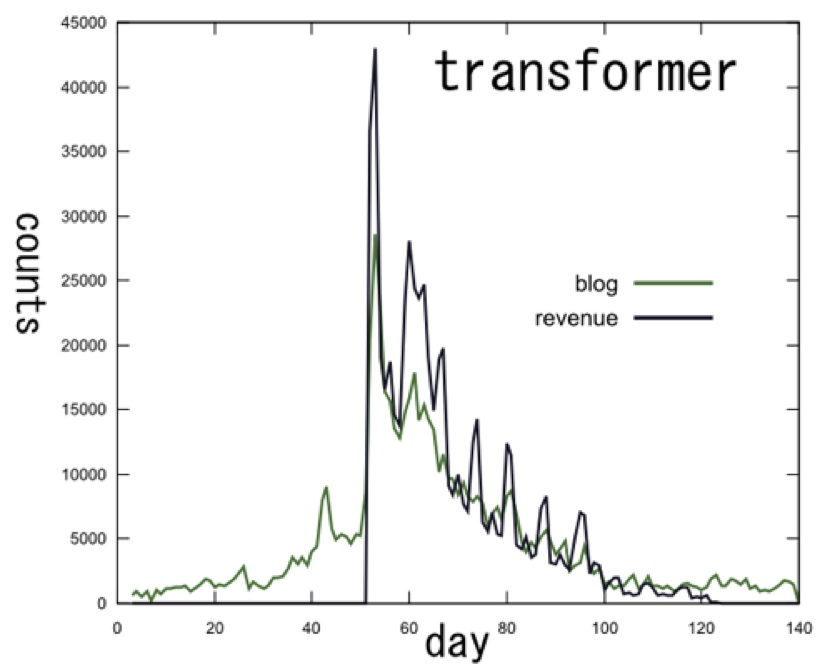}
\caption{The daily data of revenue and the blog-writing entry number for Transformer in Japanese market. The blog data are normalized to adjust the peak of the revenue at the release day.}
\label{fig:fig5}
\end{center}
\end{figure}

\section{Mathematical Model}

\subsection{Purchase intention for individual person}

Based on the observation of the movie hit phenomena in Japanes market, we present a mathematical model to explain and predict hit phenomena. First, instead of the number of audience $N(t)$, we introduce here the integrated purchase intention of individual customer, $J_i(t)$ defined as follows,

\begin{equation} 
\label{eq:eq3}
N(t) = \sum_i J_i(t)
\end{equation}

here the suffix j corresponds to individual person who has attention to the movie. Substitute (\ref{eq:eq3}) into (\ref{eq:eq1}), we obtain,

\begin{equation}
\frac{d}{dt}(\sum_i J_i(t)) = a(N_0 - \sum_i J_i(t)) = a\sum_i (1-J_i(t))
\end{equation}

Thus, we obtain the equation for $J_i(t)$ as 
\begin{equation}
\frac{dJ_i (t)}{dt} =a(1-J_i (t))
\end{equation}

when  $J_i (t)$ at$t=0$, we have the solution as

\begin{equation}
J_i(t) = 1-e^{-at}
\end{equation}

The daily purchase intention is defined from $J_i(t)$ as follows, 
\begin{equation}
\frac{dJ_i(t)}{dt} = I_i(t)
\end{equation}

Thus, we can rewrite the equation (\ref{eq:eq1}) to the equation for the purchase intention as the following way.

\begin{equation}
\frac{dI_i(t)}{dt} = -a I_i(t)
\end{equation}

This equation is the base of our mathematical model presented in the next subsection. The number of integrated audience can be calculated using the purchase intention as follows,

\begin{equation}
N(t) = \int_0^t \sum_i I_i(\tau) d\tau
\end{equation}

We extend the purchase intention to the range before the movie release date, because the customer has each purchase intention even before the movie release because of the advertisement of the movie. The purchase intention of the individual person will increase rapidly toward the release date of the movie because of the concentrated advertising campaign. Since we pointed out in the previous section that the number of the entry of the blog-writing can be considered as the quasi-revenue, it is interesting to compare the quasi-revenue and the extended purchase intention before the release of the movie.

  Since the purchase intention of the individual customer increase due to both the advertisement and the communication with other persons, we construct a mathematical model for hit phenomena as the following equation. 

\begin{equation}
\frac{dI_i(t)}{dt} = -aI_i(t) + advertisement + communication
\end{equation}

\subsection{advertisement}

Advertisement is the very important factor to increase the purchase intention of the customer in the market. Usually, the advertisement campaign is done at TV, newspaper and other medias. We consider the advertisement effect as an external force to the purchase intention as follows,

\begin{eqnarray}
\label{eq:eq12}
\frac{dI_i(t)}{dt} = -aI_i(t) + A(t) + \sum_j D_{ij} I_j(t) \nonumber \\
+ \sum_j \sum_k P_{ijk} I_j(t) I_k(t)
\end{eqnarray}

where $D_{ij}$ is the factor for the direct communication and $P_{ijk}$ is the factor for the indirect communication. Because of the term of the indirect communication, this equation is a nonlinear equation.

\subsection{Mean field approximation}

To solve the equation (\ref{eq:eq12}), we introduce here the mean field approximation for simplicity. Namely, we assume that the every person moves equally so that we can introduce the averaged value of the individual purchase intention. 

\begin{equation}
I = \frac{1}{N_p} \sum_j I_j(t)
\end{equation}

where introducing the number of potential audience $N_p$.  We obtain the direct communication term from the person who do not watch the movie as follows,

\begin{eqnarray}
\sum_j D_{ij} I_j(t) = N_p \frac{1}{N_p} \sum_j D_{ij} I_j(t)  \nonumber \\
\Rightarrow \frac{N_p -N(t)}{N_p} (N_p-N(t))D^{nn}I
\end{eqnarray}

where $D^{nn}$ is the factor of the direct communication between the persons who do not watch the movie at the time $t$. 

Similarly, we obtain the indirect communication term due to the communication between the person who do not watch the movie at the time $t$, 

\begin{eqnarray}
\sum_j \sum_k P_{ijk} I_j(t) I_k(t) = N_p \frac{1}{N_p} \sum_j N_p \frac{1}{N_p} \sum_k  \nonumber \\
\Rightarrow (\frac{N_p-N(t)}{N_p})^3 N_p^2 P^{nn} I^2 = \frac{(N_p-N(t))^3}{N_p} P^{nn} I^2
\end{eqnarray}

where $P^{nn}$ is the factor of the indirect communication between the persons who do not watch the movie at the time $t$.

  For the direct communication between the watched person and the unwatched person can be written as follows,
  
\begin{equation}
\sum_j D_{ij} = N_p \frac{1}{N_p} \sum_j D_{ij} I_j(t) \Rightarrow \frac{N(t)}{N_p} (N_p-N(t)) D^{ny} I
\end{equation}  

where $D^{ny}$ is the factor of the direct communication between the watched person and the unwatched person. For the indirect communication, we obtain more two terms corresponding to the indirect communication due to the communication between the watched persons and that between the watched person and the unwatched person as follows,

\begin{equation}
\frac{(N(t))^2 (N_p-N(t))}{N_p} P^{yy} I^2 + \frac{N(t) (N_p-N(t))^2}{N_p} P^{ny} I^2
\end{equation}

where $P^{yy}$ is the factor of the indirect communication between the watched persons and $P^{ny}$ is the factor of the indirect communication due to the communication between the watched person and the unwatched person at the time $t$. 

  Finally, we obtain the equation of the mathematical model for hit phenomena within the mean field approximation as follows,

\begin{eqnarray}
\label{eq:eq22}
\frac{I(t)}{dt} &=& -aI(t)  + A(t) + \frac{N_p -N(t)}{N_p} (N_p-N(t))D^{nn}I \nonumber \\
&& + \frac{N(t)}{N_p} (N_p-N(t)) D^{ny} I \nonumber \\   
&& +  \frac{(N_p-N(t))^3}{N_p} P^{nn} I^2 \nonumber \\
&& + \frac{(N(t))^2 (N_p-N(t))}{N_p} P^{yy} I^2 \nonumber \\
&&+ \frac{N(t) (N_p-N(t))^2}{N_p} P^{ny} I^2
\end{eqnarray}

where

\begin{equation}
\label{eq:eq23}
N(t) = N_p \int_0^t I(\tau) d\tau
\end{equation}

Thus, the equation (\ref{eq:eq22}) with (\ref{eq:eq23}) is the nonlinear integrodifferential equation. However, since the handling data is daily, the time difference is one day, we can solve the equation numerically as a difference equation.

\section{Calculated Results}

 Using the equation (\ref{eq:eq22}) and (\ref{eq:eq23}), we calculate the purchase intention for several movies where the advertisement cost presented from Dentsu is inputted into $A(t)$ with the unit of 1000 yen. The calculated results are shown in figs.6-8. The results are compared with the counts of the blog-writing entry as the quasi-revenue. We found that the agreement of the calculation with the quasi-revenue (blog) is very well. For the Da Vinci Code and the Pirates of Carribean at World's End, we have used almost same parameters except the initial value of the purchase intension. 
 
\begin{figure}[htbp]
\begin{center}
\includegraphics[width=12cm,bb=0 0 1257 1015]{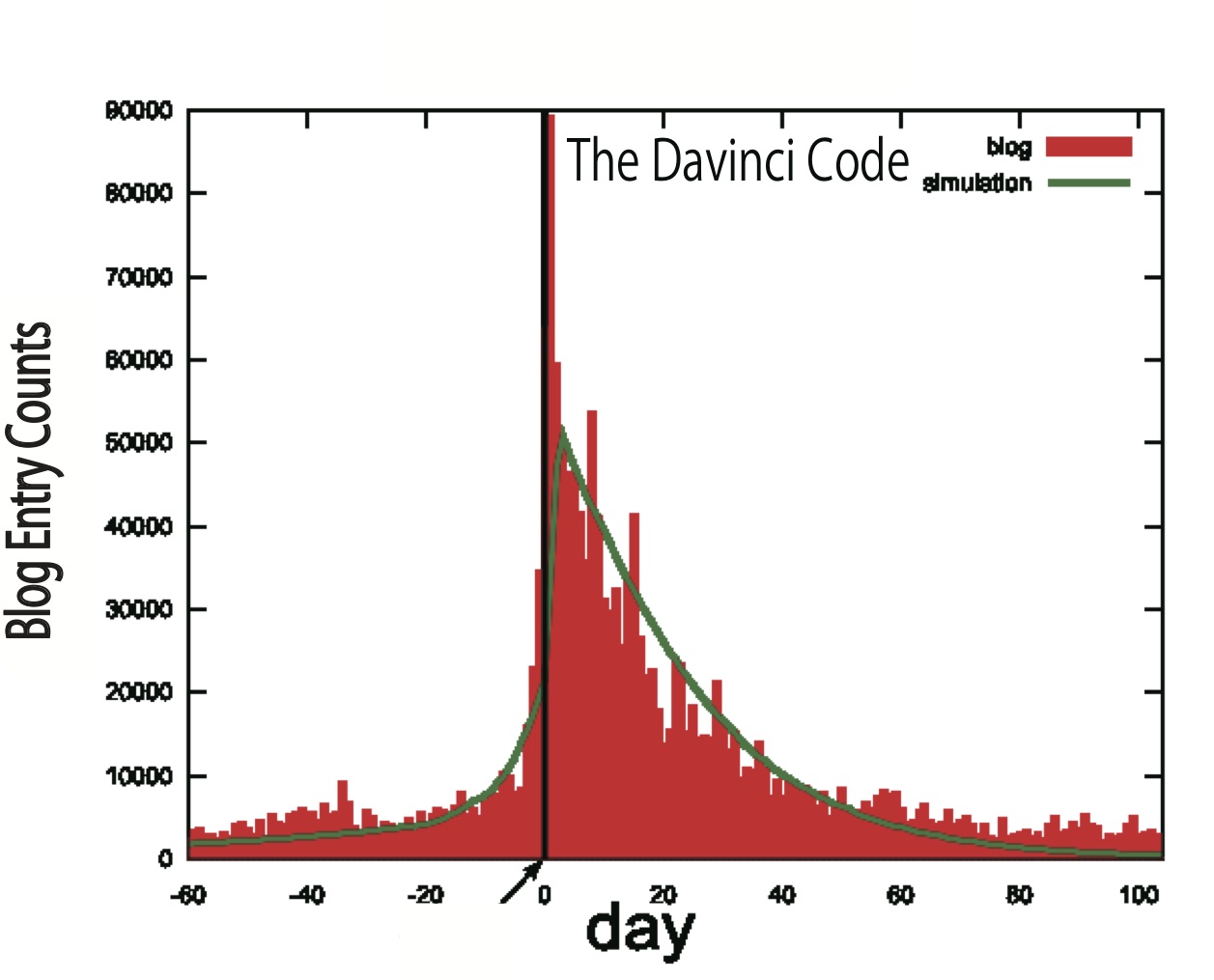}
\caption{Daily data of the calculated purchase intention (green) and the observed number of blog-writing entry(red) for the Da Vinci Code in Japanese market. .}
\label{fig:fig6}
\end{center}
\end{figure} 
 
\begin{figure}[htbp]
\begin{center}
\includegraphics[width=12cm,bb=0 0 1250 1001]{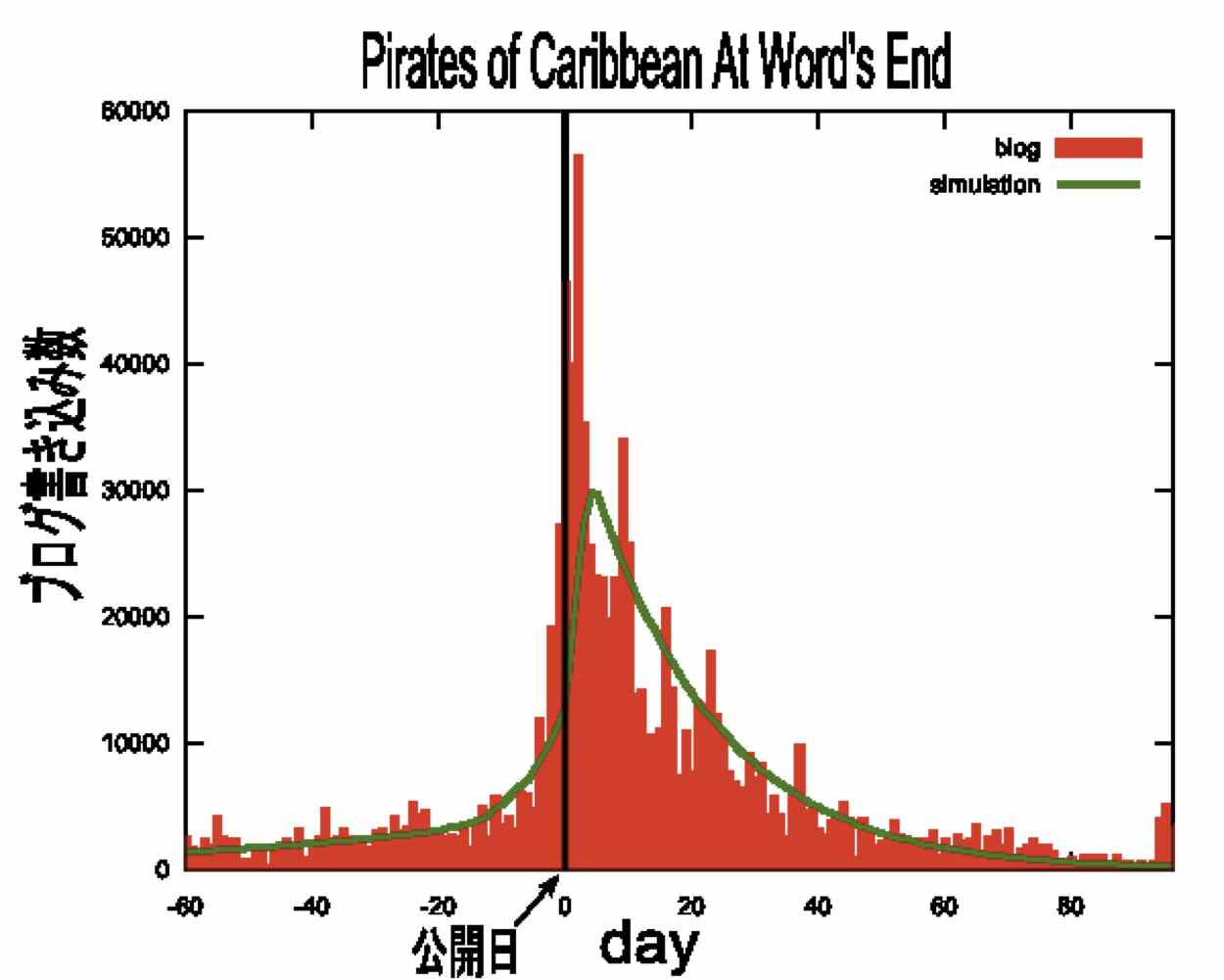}
\caption{Daily data of the calculated purchase intention (green) and the observed number of blog-writing entry(red) for Pirates of Carribean At World's End in Japanese market. .}
\label{fig:fig7}
\end{center}
\end{figure}

  Interesting results are shown in figs.6 and 7. In these figures, we present the calculation for "Always" and "Zoku Always". The two movies are the part 1 and the part 2 of the same story. Though Always is the very successful movie, the information for the movie before the opening was almost nothing. Thus, the factors of the direct and indirect communication have very large values after the opening day. For the part 2, the Zoku Always, the factors are large values even before the opening. 
    
  Finally, we show the components in the simulation of the Da Vince Code. The components shown in figure 8 correspond to the terms of eq(\ref{eq:eq22}). the results shows us that the indirect communication terms are effective for the case of the Da Vinci Code.

\begin{figure}[htbp]
\begin{center}
\includegraphics[width=12cm,bb=0 0 1250 1001]{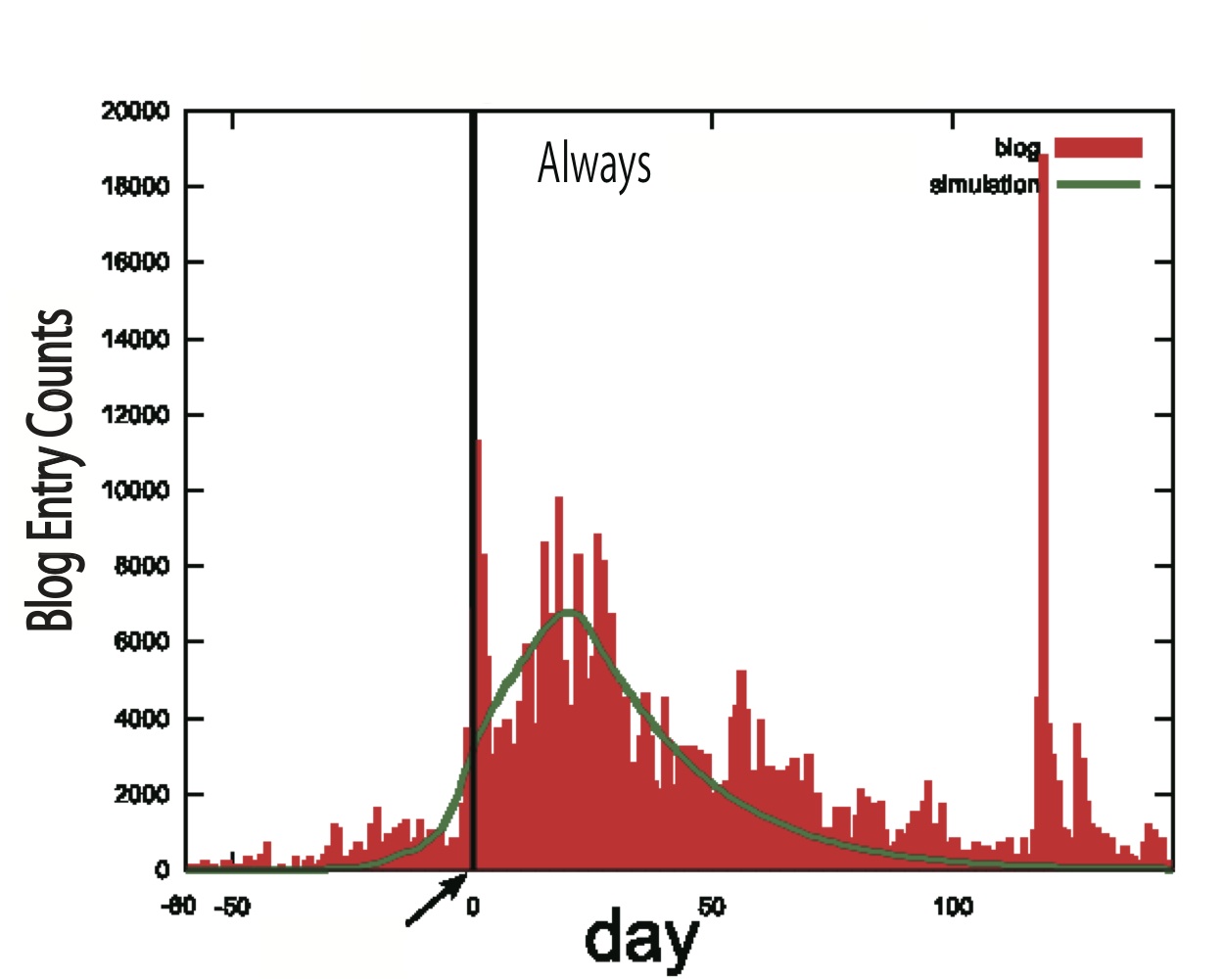}
\caption{Daily data of the calculated purchase intention (green) and the observed number of blog-writing entry(red) for "Always" in Japanese market.}
\label{fig:fig9}
\end{center}
\end{figure} 

\begin{figure}[htbp]
\begin{center}
\includegraphics[width=12cm,bb=0 0 1250 1001]{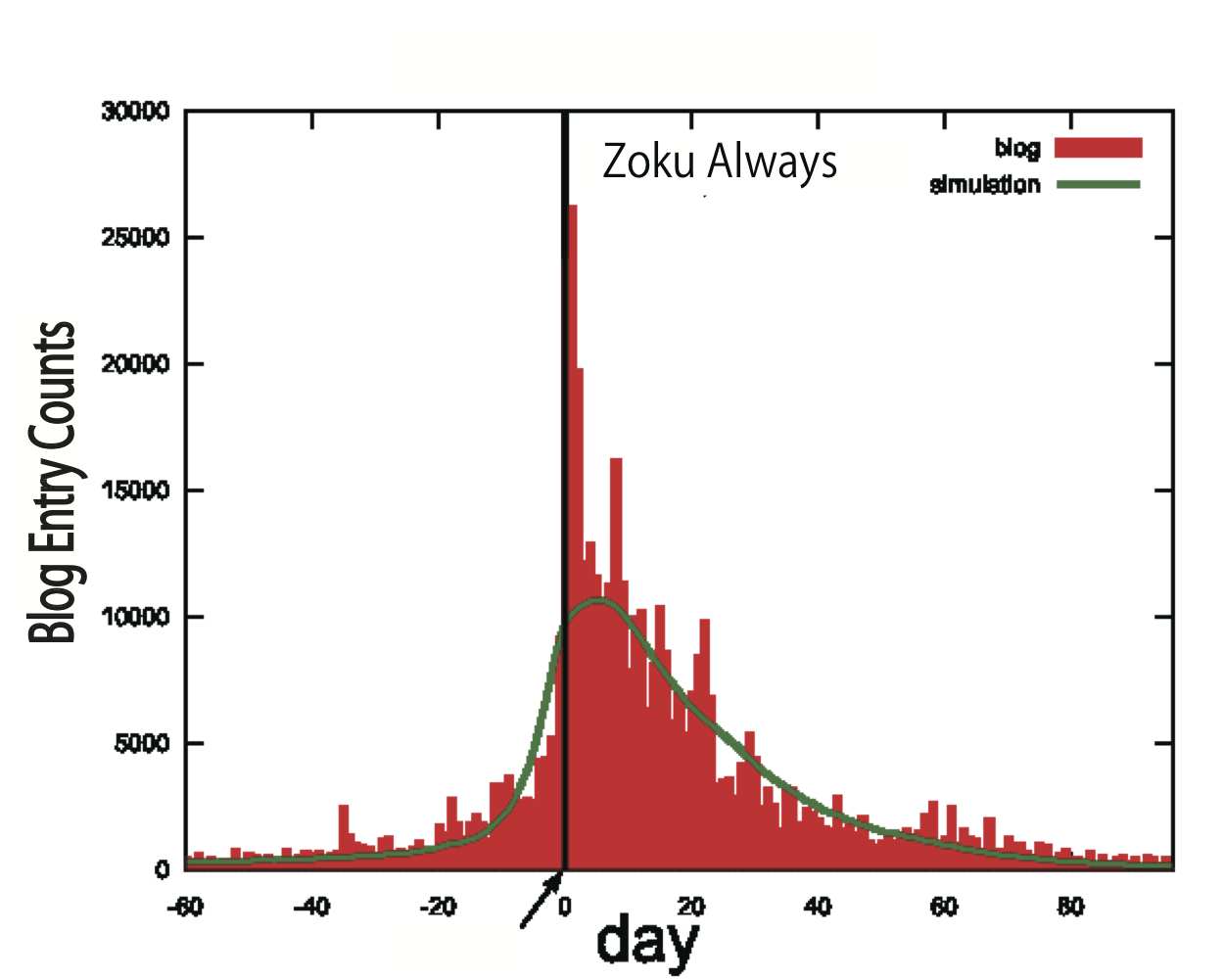}
\caption{Daily data of the calculated purchase intention (green) and the observed number of blog-writing entry(red) for "Zoku Always" in Japanese market.}
\label{fig:fig10}
\end{center}
\end{figure} 

\begin{figure}[htbp]
\begin{center}
\includegraphics[width=12cm,bb=0 0 642 482]{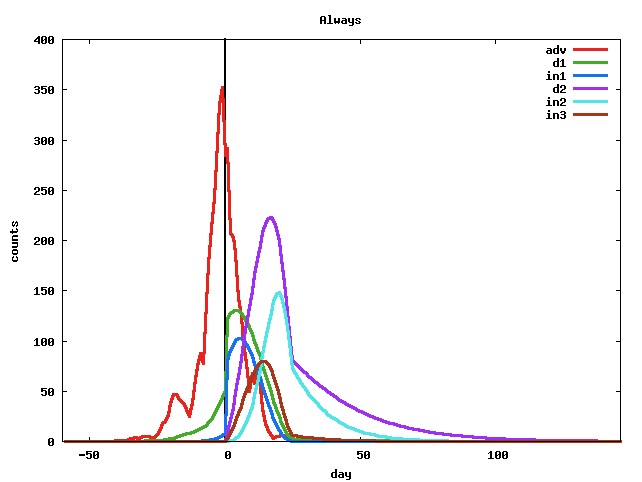}
\caption{The decoupled components of the calculation of the daily purchase intention and the daily revenue data are shown for the Da Vinci Code in Japanese market.  adv, d1, d2, in1, in2 and  in3 correpond to the terms of eq(\ref{eq:eq22})}
\label{fig:fig11}
\end{center}
\end{figure}

\section{Disucssion}

  With the agreement in figs4 and 5, we found that our model can reproduce the purchase intention very well. If we can assume that the attractions of the two movies are same, this success of the reproduction of the purchase intention means that our model can predict the revenue of the movies as a function of the advertisement cost. 
  
  The validity of the mean field approximation we introduced can be confirmed in fig.9 where we compare the present calculation for Always with the calculation of the mathematical model for hit phenomena using the scale free network model as the network of the direct communication instead of the mean field approximation. The detail of model using the scale free network will be published elsewhere\cite{Matsuda}.  The indirect communication is not included in the calculation of the scale free  network. The result shows us that the mean field approximation is farly good approximation, though the tail of the counts of blog entry can have the information of the human communication in the society.
  
\begin{figure}[htbp]
\begin{center}
\includegraphics[width=12cm,bb=0 0 978 811]{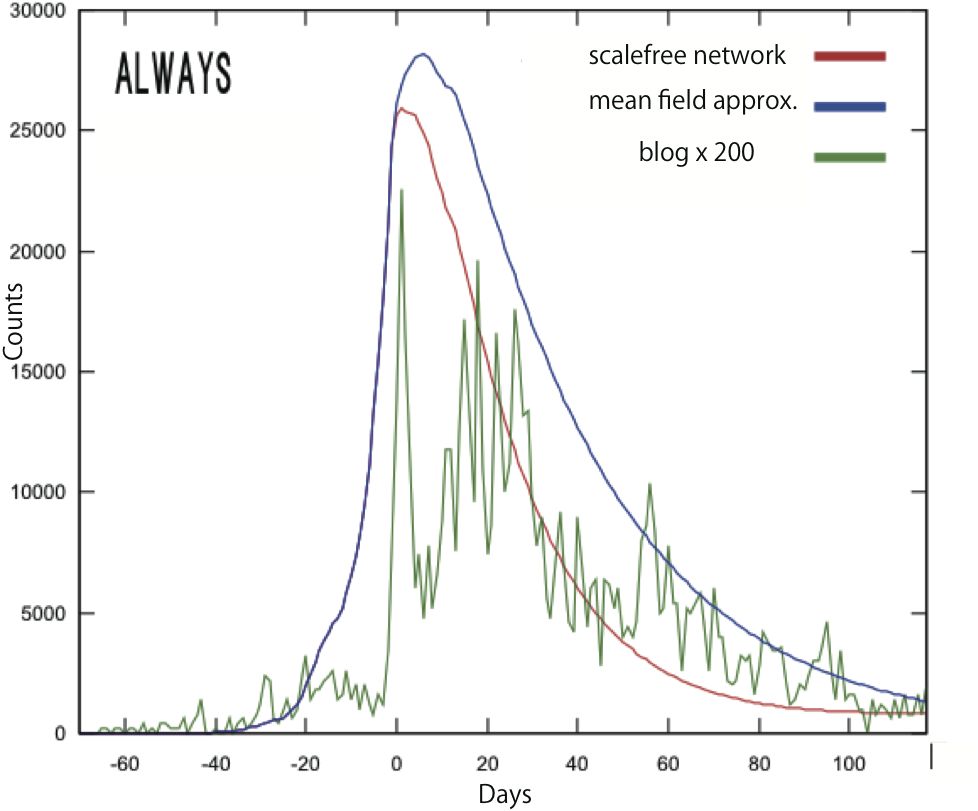}
\caption{The comparison of the calculation of the mathematical model for hit phenomena using the mean field approximation and the scale free network are shown for the Da Vinci Code in Japanese market. }
\label{fig:fig12}
\end{center}
\end{figure} 

  Therefore, we can conclude that the mathematical model presented in this paper can simulate hit phenomena at least for movie entertainment. It is very important meanings in the marketing science, because it is possible to predict whether movie will hit or not. However, we should mind that the case of the movie entertainment is very simple case as hit phenomena, because there ara no competition with other movies, in principle. No one who cannot watch the Da Vinci Code because of occupied seats tries to watch Harry Potter. Each movie can be considered almost individually.  In order to predict hit of other products like car, computer, foods or fashions, we should take into account the competition with other rival products in the same market. 
  
  Our mathematical model of hit phenomena include the indirect communication term. As we shown in ref.\cite{ishii2007}, if we neglect the indirect communication term in our model, we obtain easily the Bass model which has been known as the model of diffusion of informations by word of mouth\cite{Bass1, Bass2, Bass3}. Since the effect of the indirect communication cannot be neglect in the adjustment of the simulation with the real blog data, we found that the Bass model cannot reproduce the big hit phenomena because of the lack of the nonlinear indirect communication term.

\clearpage

\section{Conclusion}

  We found the counts of blog-writing entry is very similar to the revenue of corresponding movie. The counts of blog-writing entry can be used as quasi-revenue. The mathematical model for hit phenomena is presented including the advertisement cost and the communication effect. In the communication effect, we include both the direct communication and the indirect communication. The results calculated with the model can predict the revenue of corresponding movie very well. The conclusion presented in this paper is very useful also in marketing science.

\section*{Acknowledgements}
The advertisement cost data is presented by Dentsu Inc. The revenue data is presented from the Nikkan Kogyo Tsushinsya. The research is partially supported by the Hit Content Laboratory Inc.

%\appendix
%\section{First Appendix} %Empty argument \section{} yields `Appendix'. 
%
%\section{Second Appendix}

\end{document}